%% file: ms.tex
\pdfoutput=1
\documentclass[journal=jctcce]{achemso}
\setkeys{acs}{articletitle = true}
\SectionNumbersOn
\usepackage{amsmath}
\usepackage{xcolor}

\usepackage{boondox-cal}

\usepackage{hhline}
\usepackage{multirow}
\usepackage{array}

\def\bra#1{\langle#1 |}
\def\ket#1{| #1 \rangle}

\usepackage{enumitem}

\usepackage{caption}
\usepackage{subcaption}
\usepackage{graphicx}
\usepackage{wrapfig}

\title{Accelerating Real-Time Coupled Cluster Methods with Single-Precision Arithmetic
and Adaptive Numerical Integration}
\author{Zhe Wang}
\affiliation{Department of Chemistry, Virginia Tech, Blacksburg, VA 24061, USA}
\author{Benjamin G. Peyton}
\affiliation{Department of Chemistry, Virginia Tech, Blacksburg, VA 24061, USA}
\author{T. Daniel Crawford}
\affiliation{Department of Chemistry, Virginia Tech, Blacksburg, VA 24061, USA}
\email{crawdad@vt.edu}
\begin{document}

\input{abstract.tex}

\newpage
\input{intro.tex}

\input{theory.tex}
\input{comp.tex}

\input{results.tex}

\input{conc.tex}

\section{Acknowledgements}

This research was supported by the U.S. National Science Foundation (grant CHE-1900420). The authors are grateful to
Prof.\ Thomas B.\ Pedersen and Håkon Kristiansen of the University of Oslo for helpful discussions and to Advanced
Research Computing at Virginia Tech for providing computational resources and technical support that have contributed
to the results reported within the paper.

\bibliography{refs.bib}
\newpage
\end{document}

%% file: abstract.tex

\section*{Abstract} \label{abstract}

We explore the framework of a real-time coupled cluster method with a focus on
improving its computational efficiency.  Propagation of the wave function via the
time-dependent Schr\"odinger equation places high demands on computing resources,
particularly for high level theories such as coupled cluster with polynomial scaling.
Similar to earlier investigations of coupled cluster properties, we demonstrate that
the use of single-precision arithmetic reduces both the storage and multiplicative
costs of the real-time simulation by approximately a factor of two with no significant
impact on the resulting UV/vis absorption spectrum computed via the Fourier transform
of the time-dependent dipole moment.  Additional speedups --- of up to a factor of 14
in test simulations of water clusters --- are obtained via a straightforward GPU-based
implementation as compared to conventional CPU calculations.  We also find that
further performance optimization is accessible through sagacious selection of
numerical integration algorithms, and the adaptive methods, such as the Cash-Karp
integrator provide an effective balance between computing costs and numerical
stability.  Finally, we demonstrate that a simple mixed-step integrator based on the
conventional fourth-order Runge-Kutta approach is capable of stable propagations even
for strong external fields, provided the time step is appropriately adapted to the
duration of the laser pulse with only minimal computational overhead.

%% file: intro.tex
\section{Introduction} \label{intro}

Although quantum chemical models for the properties of stationary states have
seen great advances over the last 60 years, both in terms of accuracy and
computational efficiency, the non-equilibrium character of time-dependent
Hamiltonians (e.g., in the presence of an external, oscillating
electromagnetic field) requires approaches based on the time-dependent
Schr\"odinger equation.\cite{McCullough1969, Kosloff1988, Li2020}
While the most common techniques take a perturbational approach and treat
spectroscopic responses in the frequency domain, thereby carefully avoiding
the often-expensive time-propagation of the wave function, explicitly
time-dependent methods have a number of important advantages over their
perturbative counterparts. First, time-dependent methods allow straightforward
connections to experimental conditions, such as fine-tuning the shape,
duration, and intensity of the external fields.  Second, such methods yield
spectroscopic properties across a wide range of frequencies via Fourier
transformation of, e.g., the time-dependent electric-dipole moment, rather
than a relatively narrow window of frequencies produced by response
techniques.  Third, with careful propagation algorithms, time-dependent
methods can permit simulation of more intense external fields than
perturbation theory approaches.  Finally, the ample resources available from
modern numerical mathematics and computer science may be brought to bear to
improve the stability and efficiency of the time propagation itself.

To exploit these advantages, a wide range of real-time methods have been
explored over the last 30 years based on a variety of approximate solutions to
the electronic Schr\"odinger equation, including Hartree-Fock
(HF)\cite{Li2005}, density-functional theory (DFT),\cite{Repisky2015, Goings2018}
configuration interaction (CI),\cite{Klamroth2003,Schlegel2007} and
coupled-cluster (CC)\cite{Huber2011, Nascimento2017, Kristiansen2020}
approaches.  Among these, real-time DFT (RT-DFT) is the most widely used for
spectroscopic applications across a range of fields from biochemistry to solid
state physics where the target systems are relatively large.\cite{Sanchez2010,
Kolesov2016, Provorse2016, Bruner2016, Goings2016} The theory behind real-time
methods is continuously under development, however, and the same shortcomings
of systematic convergence and limited robustness that apply to ground-state
density-functionals also apply to RT-DFT.  This motivates researchers to
investigate higher-level methods.

Real-time coupled cluster (RT-CC) methods, in particular, can achieve
exceptionally high accuracy in many cases\cite{Bartlett2010,
Crawford2000} and have been explored in the context of real-time
simulations for some years.\cite{Huber2011, Pedersen2019, Kristiansen2020,
Pedersen2020,
Nascimento2016linear, Nascimento2017, Nascimento2019, Koulias2019, Vila2020,
Park2019, Park2021, Cooper2021} However, RT-CC approaches also suffer from
the same affliction as that of their time-independent counterparts,
\textit{viz.}, high-degree polynomial scaling with the size of the molecular
system.  For ground-state coupled cluster theory, techniques such as local
correlation, \cite{Pulay1983, Hampel1996, Neese2009, Schutz2001,Yang2012,
Russ2004, McAlexander2012, Crawford2019} fragmentation,\cite{Gordon2012,
Epifanovsky2013, Li2004, Li2009} tensor decomposition,\cite{Kinoshita2003,
Koch2003, Hohenstein2012, Schutski2017, Parrish2019, Pawlowski2019} and others
have been developed to permit applications to larger molecular systems than
conventional implementations allow.  However, while such methods are
potentially transferable to the corresponding time-dependent approaches, they
have yet to be exploited to reduce the computational cost of RT-CC.

In addition to the development of more compact representations of the
time-dependent wave function, the construction of the differential equations
that represent the time-dependency of the relevant properties, as well as the
choice of numerical integration algorithm can substantially affect the
efficiency and/or the stability of the calculation.  For example, DePrince and
Nascimento\cite{Nascimento2016,Nascimento2017} introduced left and right coupled
cluster dipole functions within the equation-of-motion (EOM-CC) framework for
calculating accurate linear absorption spectra across a wide frequency range.
They demonstrated that propagating either the left- or right-hand dipole
functions yielded the same result, thus reducing the computational cost by a
factor of two compared to propagating both left- and right-hand coupled cluster
wave functions. In 2016, Lopata\cite{Bruner2016} and coworkers accelerated
RT-DFT calculations of broadband spectra by applying Pad{\'e} approximants to
the Fourier transforms. Note that they gained a five times shorter simulation
time by obtaining rapid convergence of spectra from Pad{\'e} approximants,
while the technique has no dependence on the level of theory. In 2019, Pedersen and
Kvaal\cite{Pedersen2019} reported symplectic integrators such as Gauss-Legendre
can provide stable implementations across long propagation times even with
relatively strong external fields. In subsequent work, Pedersen, Kvaal, and
co-workers\cite{Kristiansen2020} used orbital-adaptive time-dependent coupled
cluster doubles (OATDCCD) to improve the stability of their RT-CC
implementation even when strong external fields result in a non-dominant
electronic ground state. For the application to core
excitation spectra, Bartlett and coworkers\cite{Park2019,Park2021}
compared time-independent (TI) EOM-CC and time-dependent (TD) EOM-CC --- as
well as contributions beyond the dipole approximation --- and concluded that
TD-EOM-CC can provide accurate spectra for both core and valance spectra. In
addition, in 2021, Li, DePrince, and co-workers\cite{Cooper2021} applied the short
iterative Lanczos integration to the time-dependent (TD) EOM-CC method for a
more efficient calculation of K-edge spectra. 

In this paper, we consider alternative numerical approaches aimed at reducing
the cost of RT-CC calculations.  In most quantum chemical programs,
numerical parameters are typically computed and stored using binary
representations translating to approximately 15 decimal digits of (double)
precision.  However, pioneering studies by Yasuda,\cite{Yasuda2008} Martinez and
co-workers,\cite{Ufimtsev2008,Ufimtsev2008quantum,Luehr2011} Aspuru-Guzik and
co-workers,\cite{Vogt2008} DePrince and Hammond,\cite{DePrince2011} Asadchev and Gordon\cite{Asadchev2012}, and Krylov
and co-workers\cite{Pokhilko2018} have demonstrated that single-precision
arithmetic in which the binary representation supports ca.\ seven decimal
digits, is sufficient --- and more cost effective --- for many applications.
Based on the success of this previous work, we have explored the use of
single-precision arithmetic in the context of RT-CC codes, particularly
for the simulation of linear absorption.
Additionally, we report a single-precision RT-CC implementation for which we
obtain significant further speed-up utilizing the parallel architecture of
graphical processing units (GPUs).

Finally, we explore a range of numerical integrators for solving the RT-CC
differential equations for time-dependent properties.  Generally, there are
three main types of such integrators: explicit, adaptive (or embedded), and
implicit.  Explicit integrators are so named because they take into account
only the output of the previous time step when calculating the results of the
current step, whereas adaptive integrators can adjust the step size at each
iteration with specialized algorithms to control the local error. Finally,
implicit integrators take in account the outputs of both the previous step and
the current step when determining the results for the next step, an approach
that is often more expensive due to the required iterative algorithm.
The Runge-Kutta (RK) family of integrators\cite{Butcher1996} includes all
three types and is commonly used for solving the initial value problem
associated with the Schr\"odinger equations-of-motion.  We have built a library of RK
integrators using all three types that are compatible with the RT-CC
algorithm, and here we compare their performance.  Moreover, inspired by
conventional adaptive integrators, we examine a mixed-step-size formalism to
customize the propagation on the fly, depending on whether the external field
is on or off.


%% file: theory.tex
\section{Theory} \label{theory}

The quantum mechanical description of a molecule or material subjected to an 
external time-dependent electromagnetic field requires solution of the time-dependent
Schr\"odinger equation (TDSE), which is given in atomic units as
\begin{equation}
\label{eq:TDSE} 
\hat{H}\ket{\Psi} = i\frac{d}{dt}\ket{\Psi},
\end{equation}
where the Hamiltonian,
\begin{equation}
\hat{H}(t) = \hat{H}_0 + \hat{V}(t),
\end{equation}
includes the molecular components, $\hat{H}_0$, and the time-dependent potential, $\hat{V}(t)$.  In the
time-dependent coupled-cluster framework,\cite{Nascimento2016,Pedersen2019,Crawford2019} we choose phase-isolated right- and left-hand wave functions,
respectively,
\begin{equation}
\ket{\Psi_{\rm CC}} = e^{\hat{T}(t)} \ket{0} e^{i \epsilon(t)},
\end{equation}
and 
\begin{equation}
\bra{\Psi_{\rm CC}} = \bra{0} \left(1 + \hat{\Lambda}(t)\right)
e^{-\hat{T}(t)} e^{-i \epsilon(t)}.
\end{equation}
In these expressions, $\ket{0}$ is the single-determinant reference state, and
$\hat{T}(t)$ and $\hat{\Lambda}(t)$ are second-quantized excitation and
de-excitation operators, respectively, relative to $\ket{0}$.  These operators are parametrized by 
time-dependent amplitudes that must be determined by propagating the TDSE, which, using the coupled-cluster
form of the wave function, yields right- and left-hand forms, \textit{viz.},
\begin{equation}
\bra{\mu} \bar{H} \ket{0} = i \frac{dt_\mu}{dt},
\label{Eq:derivT}
\end{equation}
and
\begin{equation}
\bra{0} \left(1 + \hat{\Lambda}\right) \left[\bar{H}, \tau_\mu\right]
\ket{0} = -i \frac{d\lambda_\mu}{dt},
\label{Eq:derivL}
\end{equation}
where the index $\mu$ denotes an excited/substituted determinant, $\tau_\mu$ is a second-quantized operator
that generates such a determinant from the reference, $\ket{0}$, and the similarity-transformed Hamiltonian,
\begin{equation}
\bar{H} = e^{-\hat{T}} \hat{H} e^{\hat{T}},
\label{Eq:simtranH}
\end{equation}
plays a central role in both the formal RT-CC equations and the algorithmic implementation.

\subsection{Propagation of the RT-CC Equations} \label{theory_1}

As mentioned in section~\ref{intro}, we propagate the CC wave function using the Runge-Kutta class of 
integrators, which are designed to solve the general initial-value problem (IVP),
\begin{equation}
\frac{dy}{dt} = f(t, y), \ \ \ \ y(t_0) = y_0,
\label{Eq:IVP}
\end{equation}
where $y$ is the unknown vector function, which is propagated in each iteration beginning from $y_0$, and $f(t,
y)$ carries the functional dependence. 
Runge-Kutta methods for solving this IVP have the general form,
\begin{equation}\label{eq:rk_y}
  y_{n+1} = y_{n} + h\sum_{i=1}^{s}b_{i}k_{ni},
\end{equation}
where
\begin{equation}\label{eq:rk_k}
  k_{ni} = f(x_{n} + c_{i}h, y_{n} + h\sum_{j=1}^{i-1}a_{ij}k_{nj}).
\end{equation}
In these expressions, $h$ is the (time) step size, and the $a_{ij}$, $b_i$,
and $c_i$ coefficients define the particular integrator.  These coefficients
may be written in a matrix form called Butcher Tableau\cite{Butcher1963} as
shown in Table.~\ref{tab:butcher}. The matrix is symmetric for explicit
integrators and asymmetric for implicit integrators, and adaptive integrators
require an additional line of coefficients.\cite{Press1992} Generally, an
additional higher-order solution can be calculated, and the difference between
the higher-order and lower-order solutions is used for adjusting the step
size. More complicated algorithms have also been developed for dynamic
simulations, higher-order differential equations, discontinuous initial-value
problems, etc.\cite{Bastani2007, Cameron1988}.
\begin{table}
    \centering
    \begin{tabular}{c|cccc}
       $c_{1}$ & $a_{11}$ & $a_{12}$ & $\cdots$ & $a_{1s}$\\ 
       $c_{2}$ & $a_{21}$ & $a_{22}$ & $\cdots$ & $a_{2s}$\\
       $\vdots$ & $\vdots$ & $ $ & $\ddots$ & $\vdots$ \\
       $c_{s}$ & $a_{s1}$ & $a_{s2}$ & $\cdots$ & $a_{s,s}$\\ \hline
        $ $ & $b_{1}$ & $b_{2}$ & $\cdots$ & $b_{s}$
\end{tabular}
\caption{Butcher tableau for Runge-Kutta integrators}
\label{tab:butcher}
\end{table}

In the RT-CC approach, we cast the time-dependent coupled-cluster left- and
right-hand amplitude expressions, Eqs.~(\ref{Eq:derivT}) and
(\ref{Eq:derivL}), into the form of Eq.~(\ref{Eq:IVP}) such that the $t_\mu$
and $\lambda_\mu$ amplitudes are collected into a single vector $y$.  Thus,
the left-hand sides of Eqs.~(\ref{Eq:derivT}) and (\ref{Eq:derivL}) provide
the specific form of $f(t, y)$.  Given that these are simply the residual
equations for the $t_\mu$ and $\lambda_\mu$ amplitudes in the time-independent
case, the initial vector, $y_0$, is naturally taken to be the solutions of the
ground-state $\hat{T}$ and $\hat{\Lambda}$ equations.  Thus, all of the same
algorithmic infrastructure used in efficient implementations of ground-state
CC methods --- spin-adaptation, intermediate factorizations, symmetry
exploitation (\textit{e.g.} using
direct-product-decomposition\cite{Stanton1991}), \textit{etc.} --- are readily applicable to the
RT-CC approach.  [NB: We choose to keep the underlying molecular orbitals from
the Hartree-Fock self-consistent field procedure from responding to the field
in order to allow direct comparison to conventional CC response and
equation-of-motion (EOM-CC) results.]

\subsection{Linear Absorption Spectra from RT-CC} \label{theory_2}

In CC theory, as with other wave-function-based methods, properties of
interest may be calculated by taking the expectation value of the
corresponding operator, though the non-Hermitian nature of the CC
similarity-transformed Hamiltonian in Eq.~(\ref{Eq:simtranH}) leads to a
generalized expectation value expression involving both the left- and
right-hand CC wave functions.  For example, the time-dependent, induced
electric dipole moment may be computed at a given time step, $t_k$, as
\begin{equation}\label{eq:dipole} \mu_{\alpha}(t_{k})=\bra{0} \left(1 +
\hat{\Lambda}(t_{k})\right) e^{-\hat{T}(t_k)}\hat{\mu}_\alpha e^{\hat{T}(t_k)}
\ket{0}(t_{k}) = \textrm{Tr}\left(\rho(t_{k})\cdot\hat{\mu}_\alpha\right)
\end{equation} where $\rho$ is the (unrelaxed) time-dependent one-particle
density matrix and $\hat{\mu}_\alpha$ is the $\alpha$-th Cartesian component
of the electric dipole operator.

If the perturbing potential is an electric field, \textit{i.e.}, 
\begin{equation}
V(t) = -\hat{\mu}_\alpha E_\beta(t),
\end{equation}
then, to a first approximation, the induced dipole moment is related to the dipole polarizability as\cite{Buckingham1967,Barron2009}
\begin{equation}
\mu_{\alpha}(t) = \alpha_{\alpha\beta}(t) \left(E_\beta(t)\right)_0,
\end{equation}
where $\alpha_{\alpha\beta}$ is the $\alpha,\beta-$th Cartesian component of the polarizability tensor, the
subscript $0$ indicates that the field is taken at the origin, and we imply the Einstein summation convention over repeated
indices.  (We note that high-intensity fields will of course induce non-linear
contributions to the induced-dipole, thus affecting the molecule's spectroscopic
response.)
The frequency-dependent dipole strength function associated with the linear absorption spectrum may
then be obtained as the imaginary component of the Fourier transform of the polarizability,
\begin{equation}
I(\omega) \propto \textrm{Im\ Tr}\left[ \boldsymbol{\alpha}(\omega) \right].
\end{equation}
In the special case that a Dirac-delta pulse is used for the shape of the electric field,
\begin{equation}
E_\beta(t) = \kappa_\beta \delta(t) \hat{n}_\beta,
\end{equation}
where $\kappa_\beta$ is the field strength and $\hat{n}_\beta$ is a unit vector in the $\beta$-th direction, then the
dipole polarizability takes a particularly simple form, \textit{viz.},
\begin{equation}
\alpha_{\alpha\beta}(\omega) = \frac{\mu_\alpha(\omega)}{\kappa_\beta}.
\end{equation}

%% file: comp.tex
\section{Computational details} \label{comp}

To test the performance of RT-CC methods, absorption spectra were calculated
from fast Fourier transform (FFT) of time-dependent induced dipole moments for
comparison with EOM-CC excitation frequencies and dipole strengths.  
We take the applied external electric field to be a Gaussian envelope,
\begin{equation}\label{eq:field}
\vec{E}(t) = - {\cal E}
e^{-\frac{1}{2}\frac{(t-\nu)^2}{\sigma^{2}}} \cos{\omega(t-\nu)} \vec{n},
\end{equation}
where the intensity ${\cal E}$, center position $\nu$ and standard
deviation $\sigma$ of the Gaussian pulse may vary for different
applications.  In addition, we take the field to be isotropic, \textit{i.e.},
\begin{equation}\label{eq:iso-field}
\vec{n} = \frac{1}{\sqrt{3}}(\hat{i} + \hat{j} + \hat{k}).
\end{equation}

All calculations were carried out using the PyCC\cite{pycc} Python-based coupled
cluster package developed in the Crawford group, which makes use of the
NumPy\cite{Harris2020} and {\tt opt\_einsum}\cite{Smith2018} packages.
PyCC provides a variety of coupled cluster methods, including CCD, CC2, CCSD,
and CCSD(T), as well as local-correlation and real-time simulations, and it
takes advantage of the ability of Python and NumPy to cast between data types
(including complex representations) automatically.  PyCC utilizes the Psi4
package\cite{Smith2020} to provide the requisite one- and two-electron integrals,
as well as the SCF molecular orbitals.  Each CPU calculation was run on a single
node with Intel's Broadwell processors, 2 x E5-2683v4 2.1GHz.  Taking advantage
of the similarity to the NumPy syntax, our GPU implementation was coded with
PyTorch 1.8.0\cite{Paszke2019} by straightforwardly substituting NumPy
functions and arrays with the corresponding PyTorch functions and tensors.  Each
GPU calculation was run on a single node with an NVIDIA P100 GPU.

Our principal test case for both the single-precision calculations and the tests
of integrators is the series of water clusters, (H$_2$O)$_n$ up to $n=4$, using
the coordinates provided by Pokhilko \textit{et al.}\cite{Pokhilko2018} All the
calculations were carried out at the coupled cluster singles and doubles level
(CCSD) with the correlation-consistent double-zeta (cc-pVDZ) basis
set.\cite{Dunning1989}  All calculations kept the $1s$ core orbitals on the oxygen
atoms frozen.  For comparison to conventional linear response results, we also
carried out EOM-CCSD/cc-pVDZ excitation-energy calculations, which are included
as stick spectra.  These results were obtained using the Psi4 code\cite{Smith2020}
with the same frozen-core approximation.

%% file: results.tex
\section{Results and Discussion} \label{results}

\subsection{Single-Precision RT-CC}\label{result:precision}

For the last 35 years (and updated most recently in 2019),the IEEE 754
standard\cite{IEEE-754-2019} has defined ``interchange and arithmetic formats
and methods for binary and decimal floating-point arithmetic in computer
programming environments,'' {\em i.e.}\ the representation and mathematical
operations of single- and double-precision numbers, among others.  In this
standard, each floating-point number is stored with three components: a sign, a
significand/coefficient, and an exponent.  For single precision, the 23 explicit
bits of the significand (plus an implicit bit for normal numbers) yields
$\log_{10}(2^{24}) \approx 7.22$ decimal digits, whereas double precision
yields $\log_{10}(2^{53}) \approx 15.95$ decimal digits.  As mentioned in
section~\ref{intro}, the development of quantum chemical methods that take
advantage of the efficiency of single- and mixed-precision arithmetic --- both
in storage and in computing time --- have seen considerable advances in recent
years.  For size-extensive properties, such as the total electronic energy,
Ufimtsev and Mart\' inez\cite{Ufimtsev2008,Ufimtsev2008quantum} demonstrated that
purely single-precision arthimetic and storage quickly becomes inadequate as the
size of the molecular system increases.  Similar observations were reported by
Asadchev and Gordon\cite{Asadchev2012} in their mixed- and high-precision
implementation of the Rys quadrature for the evaluation of two-electron
repulsion integrals, and by Tornai and co-workers in their development of a
high-performance, dynamic integral-evaluation program.\cite{Tornai2019}
Pokhilko, Epifanovsky, and Krylov\cite{Pokhilko2018} also reported a novel
single- and mixed-precision implementation of CC and equation-of-motion CC
methods in 2018, in which solution iterations of the relevant CC equations in
single precision were adequate for most applications, and a limited number of
``cleanup'' iterations in double-precision could recover higher accuracy.  Thus,
for such cases, a mixed-precision approach wherein some steps of the quantum
chemical calculation are carried out in single precision and others in higher
precision becomes essential.  

In this work we focus on simulations of linear UV/vis absorption spectra using
the approach described earlier.  If single-precision arithmetic is adequate, the
computational cost can be reduced to nearly half for both small and large
systems, and the error will not accumulate as the system size increases,
assuming neither the dipole moment nor the relevant electronic-excitation
domains are extensive.  To test this assumption, we computed the time-dependent
dipole moments of the water molecule at a series time points from a
double-precision RT-CCSD/cc-pVDZ calculation and corrupted the data by adding
random noise at several magnitude cutoffs.  In this initial test, we carried out
the time propagation for 300 a.u.\ using a Gaussian envelope with a field
strength ${\cal E} = 0.01$ a.u., center $\nu = 0.05$ a.u., width $\sigma = 0.01$
a.u., and a time step $h = 0.01$ a.u.

\begin{figure}[H]
    \centering
    \includegraphics[angle=0, scale=0.4]{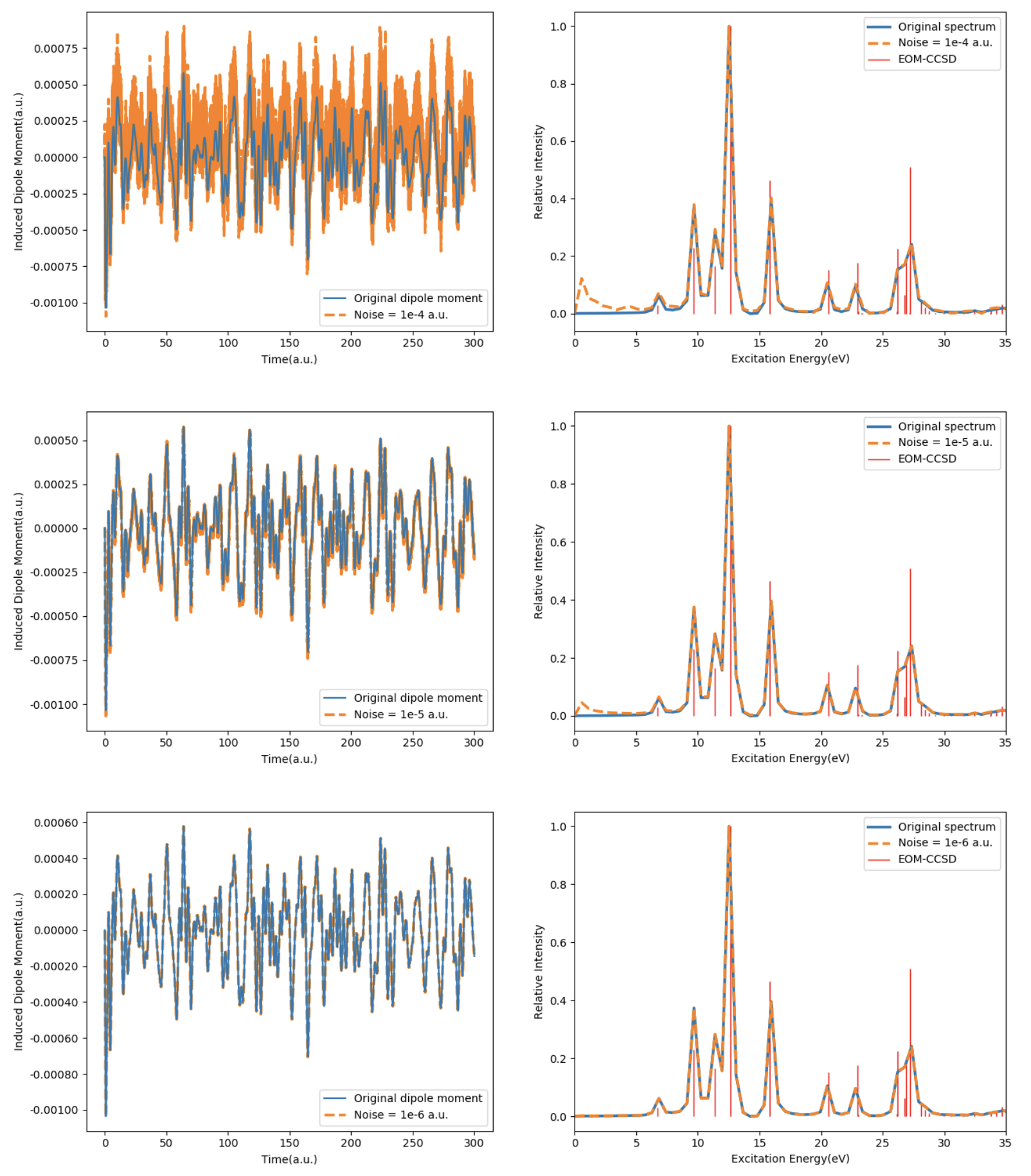}
    \caption{RT-CCSD/cc-pVDZ time-dependent induced electric dipole moments (left-hand
column) for a water molecule in the presence of an external electric field and
the corresponding linear absorption spectrum (right-hand column) with and
without random noise of varying magnitudes. Corresponding EOM-CCSD/cc-pVDZ 
transitions are included as stick-spectra for comparison.}
    \label{fig:sp-noise}
\end{figure}

As shown in Fig.~\ref{fig:sp-noise}, the spectrum begins to deviate from
the original dipole trajectory only with random noise added starting at a magnitude
of $10^{-5}$ and greater.  In such cases, the spectra manifest the appearance of
the noise typically in the low-frequency region, as can be seen in the upper two
spectra between 0-5 eV.  Smaller noise cutoffs yield spectra that are
indistinguishable from their noise-free counterpart.  This is consistent with
the expectation based on the IEEE definition of floating-point arithmetic,
namely that single-precision can retain accuracy to roughly $10^{-7}$. 

To compare the double-precision and single-precision arithmetic directly, we
calculated RT-CCSD/cc-pVDZ dipole trajectories and the corresponding linear
absorption spectra using both representations for the series of (H$_2$O)$_n$
clusters with $n=1-4$. For these simulations, the explicit integration was
carried out using the Runge-Kutta 4th order integrator (RK4) with a step size
$h=0.01$ a.u.  The external field was chosen to be a Gaussian envelope defined
in Eq.~(\ref{eq:field}) with ${\cal E} = 0.01$ a.u., $\nu=0.05$ a.u., and
$\sigma=0.01$ a.u. (a narrow pulse). The results are aligned with the numerical
experiments above, with no discernible difference in the spectra after lowering
the arithmetic precision to single-precision. All the spectra are also compared
with EOM-CCSD/cc-pVDZ calculations: we include 40 EOM-CC roots in each of the
spectra, all of which are well-aligned with the associated RT-CC peaks. From this perspective,
the computation time and the required size of memory can both be reduced by
ca.\ a factor of two for the calculation of the spectra, as previously observed
for electronic energies and other components of quantum chemical
calculations.\cite{Ufimtsev2008,Ufimtsev2008quantum,Asadchev2012,Tornai2019,Pokhilko2018}
We note, however, that single-precision arithmetic may not be practical for certain numerically sensitive calculations,
\textit{e.g.}, using higher-order numerical differentiation to extract linear and nonlinear response functions as reported
by Ding \textit{et al.}\cite{Ding2013}
(In addition, we note that these spectra are not intended to reproduce vapor-phase experimental
measurements, only to test the validity of the single-precision arithmetic
approximation.  Thus, any transitions appearing above the physical ionization
limit are not physically realistic and are merely an artifact of the use of a
finite basis set without representation of continuum states.)

\begin{figure}[H]
     \centering
     \begin{subfigure}[H]{0.475\textwidth}
         \centering
         \includegraphics[width=\textwidth]{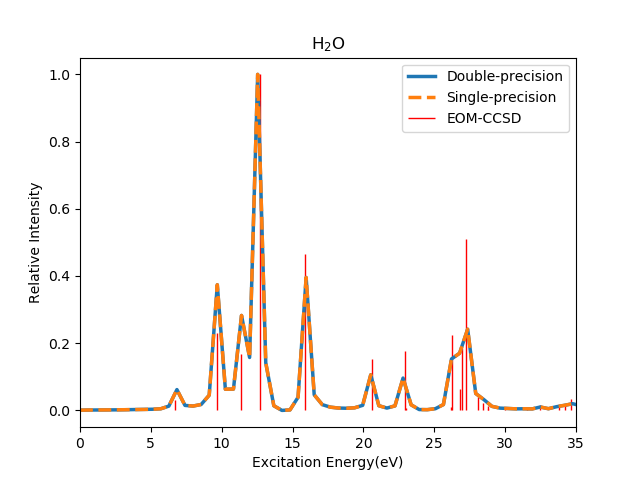}
         \label{fig:sp-monomer}
     \end{subfigure}
     \hfill
     \begin{subfigure}[H]{0.475\textwidth}
         \centering
         \includegraphics[width=\textwidth]{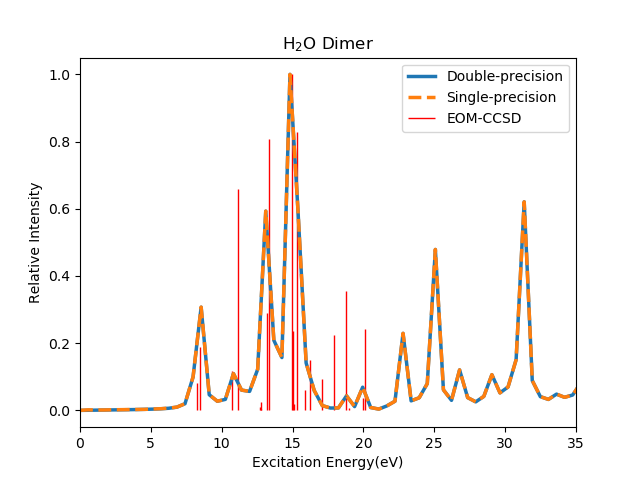}
         \label{fig:sp-dimer}
     \end{subfigure}
     \vfill
     \begin{subfigure}[H]{0.475\textwidth}
         \centering
         \includegraphics[width=\textwidth]{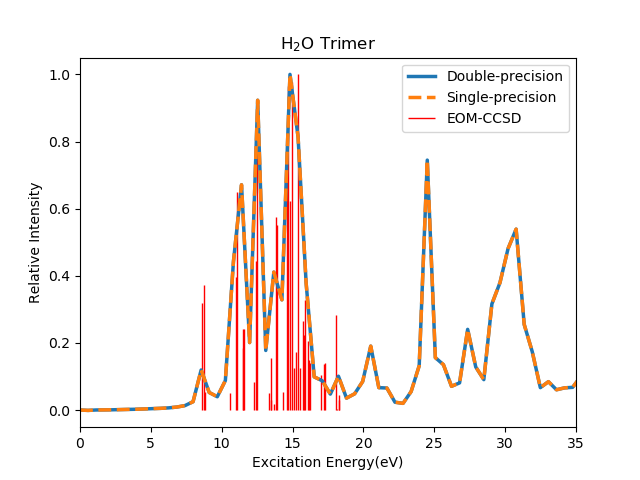}
         \label{fig:sp-trimer}
     \end{subfigure}
     \hfill
     \begin{subfigure}[H]{0.475\textwidth}
         \centering
         \includegraphics[width=\textwidth]{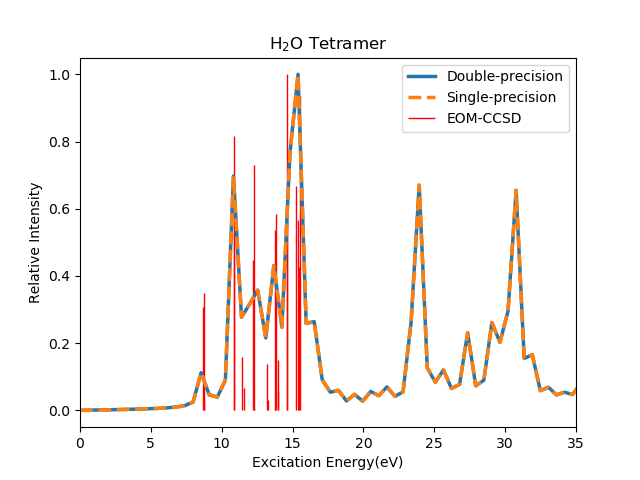}
         \label{fig:sp-tetramer}
     \end{subfigure}
     \caption{Linear UV/vis absorption spectra of (H$_{2}$O)$_n$ clusters for
$n=1-4$ calculated at the RT-CCSD/cc-pVDZ level of theory in both double- and
single-precision arithmetic. Time propagation was carried out for 300 a.u. in
the presence of a weak electric field represented by a narrow Gaussian pulse.
Corresponding EOM-CCSD/cc-pVDZ transitions are included as stick-spectra for
comparison.}
     \label{fig:sp-spectrum}
\end{figure}

Inspired by many parallel implementations of CC methods for distributed memory
architectures on CPUs\cite{Olson2007, Solomonik2014, Janowski2007,
Anisimov2014}, corresponding GPU implementations have become desirable in order
to take advantage of heterogenous artchitectures on modern high-performance
computing systems.  While early GPU hardware was designed for accelerating image
processing with an emphasis on mostly single- or low-precision floating point
operations for quick memory access when higher accuracy is not required, over
the past several years, GPUs have been more extensively used for scientific
research.  In addition to the development of GPUs with robust performance for
double-precision arithmetic, numerous software toolkits have also emerged such
as the Computer Unified Device Architecture (CUDA)\cite{cuda}, the Open
Computing Language (OpenCL),\cite{Stone2010} and a variety machine-learning
packages that support GPUs such as TensorFlow,\cite{tensorflow2015}
PyTorch,\cite{Paszke2019} and others, all of which lower the barriers to a wide
range of scientific applications that can take advantage of modern GPU
performance.  We note that, even though the performance of double-precision
calculations on GPUs is already relatively robust, single-precision arithmetic
is still preferable if it provides neglible errors relative to double-precision
results due to the substantial improvement in computational speed and memory
usage.  

For the RT-CC methods explored in this work, we have therefore developed a
GPU-capable implementation within the PyCC code using the PyTorch
package\cite{Paszke2019} based on the conventional CPU version described in
section~\ref{comp}.  In the PyCC implementation, all one-electron quantities
such as the Fock matrix (including the external field), $\hat{T}_1$, and
$\hat{T}_2$ amplitudes are loaded onto the GPU at the beginning of either each
iteration of the time-independent wave function or each computation of the
residuals in Eqs.~(\ref{Eq:derivT}) and (\ref{Eq:derivL}) during the RT-CC
propagation.  As each term in the CC equations is evaluated, the necessary
subblock of the two-electron repulsion integrals is loaded onto the GPU, and the
required tensor contraction is carried out using the usual {\tt opt\_einsum}
function.  Other two- and four-index intermediates formed from the
similarity-transformed Hamiltonian (\textit{e.g.}, the $W_{mbej}$ or $W_{amef}$
intermediates), are retained on the GPU as they are created for later use in the
same iteration/time-step, and deleted once the current residual is complete.
The advantage of this approach is that it provides straightforward access to
local GPU hardware without significant modification of the NumPy-based
tensor-contraction code already in place.  Of course, this Python-based
implementation does not represent the full performance of a production-level
code and is necessarily limited in terms of the size of the molecular system it
can treat, but it does provide a valuable estimate of the minimum expected
speed-up one can obtain for a more highly optimized alogrithm.

Table~\ref{tab:gpu-cpu} provides a comparison of RT-CCSD/cc-pVDZ timings for
double-precision (dp) and single-precision (sp) on CPUs and GPUs for our set of
example water clusters, using the same Gaussian envelope parameters are for
previous computations.  The first three columns report the number of seconds
required for each time step of the RT-CC simulation averaged over a 300 a.u.
propagation (\textit{i.e.}, for a step size of $h=0.01$ a.u., averaged over
30,000 time steps).  As the size of the molecular system increases from monomer
to tetramer, the computational cost per iteration for a CPU-dp calculation
increases by approximately a factor of ca.\ $4^{4.85}$, whereas for a GPU-dp
calculation, the increase is a factor of ca.\ $4^{3.18}$, and for a GPU-sp
calculation, this falls to $4^{2.88}$.  While these are clearly less than the
formal ${\cal O}(4^6)$ scaling expected from CCSD, all of these implementations
would eventually reach that limit for larger systems, because the CPU/GPU and
dp/sp improvements only affect the prefactor, and not the exponent.
Nevertheless, the use of the GPU coupled with single-precision arithmetic
clearly offers substantial advantages.  This improvement is also clearly seen in
the data reported in the final two columns of Table~\ref{tab:gpu-cpu}, where the
single-precision code yields roughly the expected factor of two speed-up over
its double-precision counterpart (with both calculations taking place on the
GPU), and the GPU offers up to a factor of 14 speed-up over the CPU when both
operate in double-precision (or better when the GPU operates in single-precision
mode).  Larger molecular systems should offer even greater improvement, with the
proviso that the memory limits of the GPU will eventually produce a performance
pleateu.
\begin{table} 
    \centering
    \begin{tabular}{c|ccccc}
       \textrm{Water Cluster} & $t_\textrm{CPU-dp}$ &  $t_\textrm{GPU-dp}$ &
$t_\textrm{GPU-sp}$ & $\frac{t_\textrm{CPU-dp}}{t_\textrm{GPU-dp}}$ &
$\frac{t_\textrm{GPU-dp}}{t_\textrm{GPU-sp}}$\\ \hline
       \textrm{Monomer} & 0.17217 & 0.14330 & 0.13253 & 1.2015 & 1.0813\\ 
       \textrm{Dimer} & 3.4705 & 0.60738 & 0.40496 & 5.7139 & 1.4999\\
       \textrm{Trimer} & 32.729 & 3.4910 & 1.7264 & 9.3752 & 2.0221 \\
       \textrm{Tetramer} & 167.43 & 11.727 & 7.2215 & 14.277 & 1.6239\\     
    \end{tabular}
    \caption{Performance comparison of conventional RT-CC/cc-pVDZ calculations
for water clusters using double-precision on the CPU (CPU-dp), double-precision
on the GPU (GPU-dp), and single-precision on the GPU (GPU-dp). Timings (first
three columns) are given in seconds as per-iteration averages over a 300 a.u.
propagation wth $h=0.01$ a.u.  The final two columns are speed-ups,
\textit{i.e.} ratios of timings for each case.}
    \label{tab:gpu-cpu}
\end{table}
        
\subsection{Comparison of Numerical Integrators for RT-CC} \label{result:integrator}

As discussed in section~\ref{theory}, the family of Runge-Kutta methods is the
most widely used class of numerical integrators for IVPs such as the TDSE, not
only because their implementation is straightforward --- \textit{e.g.}, for
RT-CC methods they are easily adapted to accept function vectors taken from the
left- and right-hand wave function residuals in Eqs.~(\ref{Eq:derivT}) and
(\ref{Eq:derivL}) --- but also because of their relatively robust performance
for a range of applications.  The classic Runge-Kutta 4th order integrator
(RK4), for example, is one of the most commonly used algorithms for scientific
problems because it averages each time step into four simple stages and yields a
small truncation error of order $h^{5}$, where $h$ is the step size,
\begin{equation}\label{eq:rk4truncation}
y_{n+1} = y_{n} + \frac{1}{6}(k_{1}+2k_{2}+2k_{3}+k_{4}) + \textit{O}(h^{5}).
\end{equation}

Such explicit integrators typically yield stable propagations of real-time
methods, provided sufficient care is taken in choosing the step size, which is
critical not only for the integration to be numerically stable, accurate, and
efficient, but also cost effective: larger step sizes reduce the computational
expense of the simulation, but they can also result in failure of the
propagation due to a non-convergent time series.  Rather than running sets of
numerical experiments to find the largest, reasonable step size that can provide
accurate results for every application, adaptive integrators\cite{Press1992}
were designed to balance the required stability with the least computational
cost by exerting algorithmic control over step size  within the existing process of
explicit integrators.  For example, Fehlberg\cite{Fehlberg1968} discovered that,
by carrying out six function evaluations per time step and using different
combinations of the six resulting intermediates to formulate a fifth-order
solution and a fourth-order solution, a fourth-order method can be derived with
step size control.  In 1990, Cash and Karp\cite{Cash1990} found another
combination of Fehlberg's coefficients that yields an even more efficient
method,
\begin{equation}\label{eq:ck-k}
\begin{aligned}
k_{1} &= f(t_{n}, y_{n})\\
k_{2} &= f\left(t_{n}+\frac{1}{5}h, y_{n}+\frac{1}{5}k_{1}h\right)\\
k_{3} &= f\left(t_{n}+\frac{3}{10}h,
y_{n}+h\left(\frac{3}{40}k_{1}+\frac{9}{40}k_{2}\right)\right)\\
k_{4} &= f\left(t_{n}+h,
y_{n}+h\left(\frac{3}{10}k_{1}+\frac{9}{10}k_{2}+\frac{6}{5}k_{3}\right)\right)\\
k_{5} &= f\left(t_{n}+\frac{3}{5}h,
y_{n}+h\left(\frac{-11}{54}k_{1}+\frac{5}{2}k_{2}-\frac{70}{27}k_{3}+\frac{35}{27}k_{4}\right)\right)\\
k_{6} &= f\left(t_{n}+\frac{7}{8}h,
y_{n}+h\left(\frac{1631}{55296}k_{1}+\frac{175}{512}k_{2}-\frac{575}{13824}k_{3}+\frac{44275}{110592}k_{4}+\frac{253}{4096}k_{5}\right)\right),
\end{aligned}
\end{equation}
with the resulting coupled time steps being,
\begin{equation}\label{eq:ck-y}
\begin{aligned}
y_{1} &=
y_{n}+h\left(\frac{37}{378}k_{1}+\frac{250}{621}k_{3}+\frac{125}{594}k_{4}+\frac{512}{1771}k_{6}\right)\\
y_{2} &=
y_{n}+h\left(\frac{2825}{27648}k_{1}+\frac{18575}{48384}k_{3}+\frac{13525}{55296}k_{4}+\frac{277}{14336}k_{5}+\frac{1}{4}k_{6}\right)
\end{aligned}
\end{equation}
where $y_{1}$ is a fourth-order solution embedded with $y_{2}$, which is a
fifth-order solution. Therefore, with $\Delta=|y_{2}-y_{1}|$ taken to be an
error estimate for the current time step of order $h^{5}$, a desired accuracy of
$\epsilon$ yields a formula for adjusting the step-size on the fly,
\textit{viz.},
\begin{equation}\label{eq:ck-h}
\frac{h_{new}}{h}=\left(\frac{\epsilon}{\Delta}\right)^{1/5}.
\end{equation}
Thus, if $\Delta$ is smaller than $\epsilon$, $h$ will be increased for the next
step, but if $\Delta$ is larger than $\epsilon$, $h$ will be reduced and the
current step must be repeated until the required accuracy is reached.
Furthermore, if $h$ is reduced and used for the current step again, the error
will have an implicit scaling of $h$, and the exponent in Eq.~(\ref{eq:ck-h})
must be shifted from $\frac{1}{5}$ to $\frac{1}{4}$. The size control of final
step is given as,
\begin{equation}\label{eq:ck-h-final}
\begin{aligned}
h_{new} &= 0.84 h \left(\frac{\epsilon}{\Delta}\right)^{1/5}   \hspace{0.5cm}&
\textrm{for} \hspace{0.1cm}|\Delta| \leq \epsilon \\
h_{new} &= 0.84 h \left(\frac{\epsilon}{\Delta}\right)^{1/4}   \hspace{0.5cm}& 
\textrm{for} \hspace{0.1cm}|\Delta| > \epsilon 
\end{aligned}
\end{equation}
where the coefficient $0.84$ is a ``safety factor'' because the error
estimates are not exact.  Thus, the computational cost of the adaptive
integrator is optimized under a predetermined desired accuracy. If the values at
consecutive time steps change rapidly, a small step size will naturally be used;
if the values vary only slightly, larger step sizes will be sufficient for a
stable propagation. 

In our RT-CC calculations, the time-dependent cluster amplitudes change rapidly
when the external field is on at the beginning of the simulation and gradually
stabilize after the field is turned off.  With this in mind, we tested the
adaptive Cash-Karp (CK) integrator described above for the RT-CCSD simulation
of the absorption spectrum of a single water molecule using the same external
field as in the calculations in section \ref{result:precision}. Additionally,
with the results shown in section \ref{result:precision}, all the calculations
are run in single-precision for the efficiency. 

\begin{figure}[H]
    \centering
    \includegraphics[angle=0, scale=0.3]{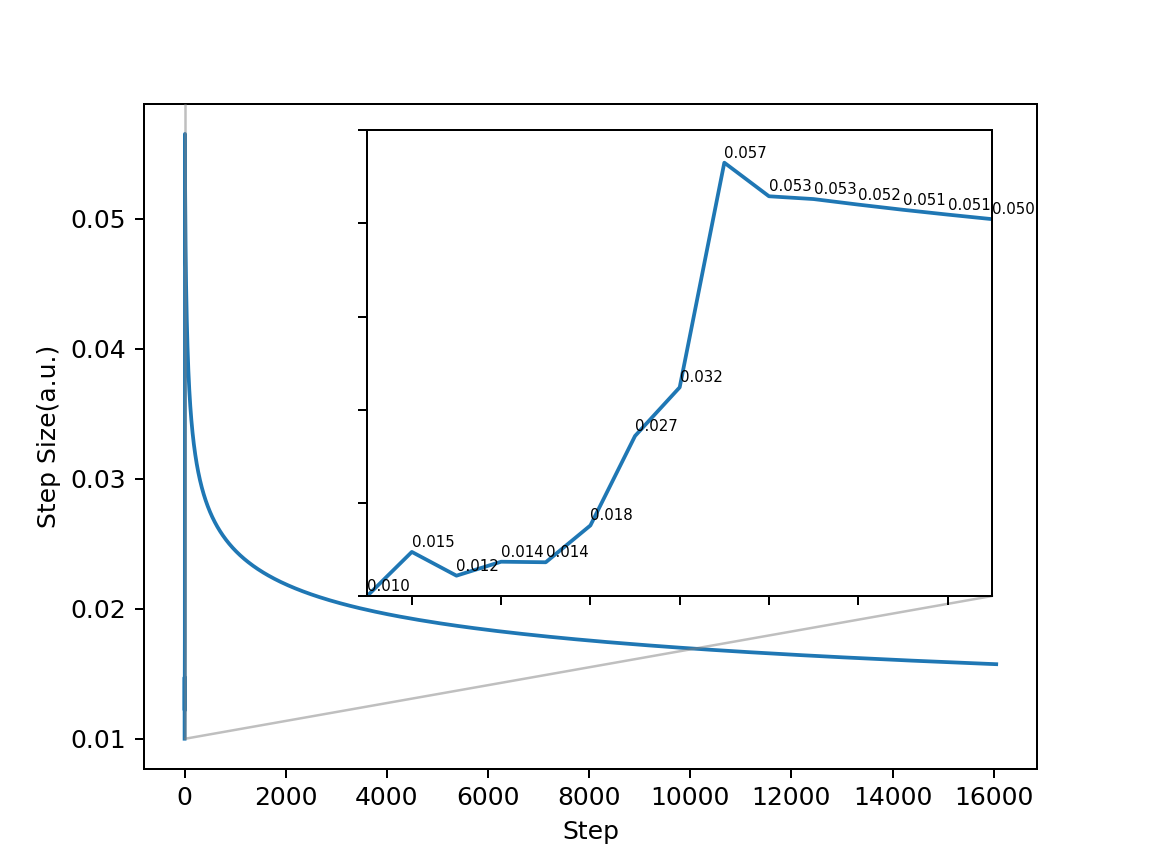}
    \caption{The adjusted step size at each time step in the RT-CCSD/cc-pVDZ
calculation for H$_2$O using the adaptive Cash-Karp integrator over 300 a.u.\
propagation. The first 15 time steps are zoomed in to focus on the detailed change.}
    \label{fig:ck-step}
\end{figure}

\begin{figure}[htbp]
    \centering
    \begin{subfigure}[H]{0.475\textwidth}
        \centering
        \includegraphics[width=\textwidth]{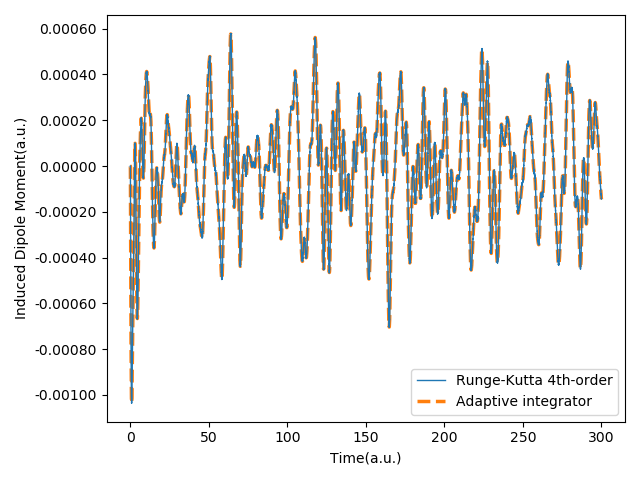}
    \end{subfigure}
    \hfill
    \begin{subfigure}[H]{0.475\textwidth}
        \centering
        \includegraphics[width=\textwidth]{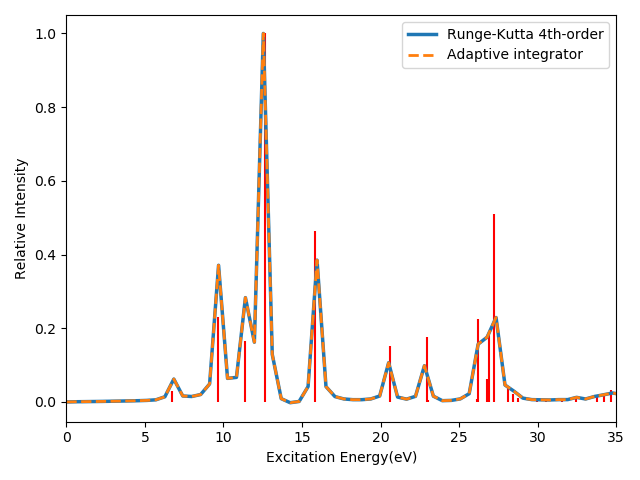}
    \end{subfigure}
    \caption{Comparison of the time-dependent induced dipole moment and the
corresponding linear absorption spectrum from the RT-CCSD/cc-pVDZ simulation of
H$_2$O using RK4 and CK integrators. EOM-CCSD/cc-pVDZ transition energies are
depicted as stick-spectra for reference.}
    \label{fig:ck-results}
\end{figure}

Fig.~\ref{fig:ck-step} reports the variation in the step size at each iteration
of the propagation using an initial step size of $h=0.01$ a.u.  From
Eq.~(\ref{eq:field}), if $t=0.01$ a.u.\ the field strength is only $3.35\times
10^{-6}$ a.u., and the local error $\Delta$, is expected to be smaller than
$\epsilon$ according to Eq.~(\ref{eq:ck-h-final}), leading to an increase in
$h$ to 0.015 a.u.  At the second time step when $t=0.025$ a.u., the
corresponding field strength is $4.39\times 10^{-4}$ a.u., which gets closer to
its peak of 0.01 a.u.  At this point in the simulation, $\Delta$ is large, and
the CK algorithm automatically reduces the step size to $h=0.012$ a.u.  At the
third and fourth time steps, $h$ slightly increases to 0.014 a.u., because,
even though the field is still on, $h=0.012$ a.u. is small enough to keep
$\Delta$ smaller than $\epsilon$. After the fifth time step, $h$ is increased
to 0.018 a.u. and further 0.027 a.u. when $t=0.064 ~\textrm{to}~ 0.082$ a.u.,
because, during this point in the propagation, the field strength has already
begun to decrease due to the brevity of the pulse.  When the propagation
reaches $t=0.109$ a.u., the magnitude of the field strength falls back to
$10^{-10}$, $\Delta$ is small when $h=0.027$ a.u.\ is tested, and thus $h$
increases to 0.032 a.u.\ at the seventh time step and 0.057 a.u.\ at the eighth
time step. Starting from the ninth time step when $t=0.198$ a.u., although the
external field is essentially zero for the remainder of the propagation, the
algorithm still converges to a step size of $h=0.02$ a.u. due to the continued
oscillation of the amplitudes instigated by the pulse. Since the step size varies 
throughout the propagation, the dipole moments are not calculated at equally 
spaced time points, typical Fourier transform will not work, instead, we pre-process 
our data by interpolating the data points to an evenly spaced grid. Ultimately, after a 300
a.u.\ propagation, the adaptive CK integrator yields an overall speedup of 1.32
relative to the explicit RK4 integrator, yet, as shown in
Fig.~\ref{fig:ck-results}, the final absorption spectra obtained from each
algorithm exhibit no significant differences.  Thus, adaptive integrators such
as the CK algorithm can provide an automated approach to systematically
optimizing the step size depending on the system.

\begin{figure}[H]
    \centering
    \includegraphics[angle=0, scale=0.3]{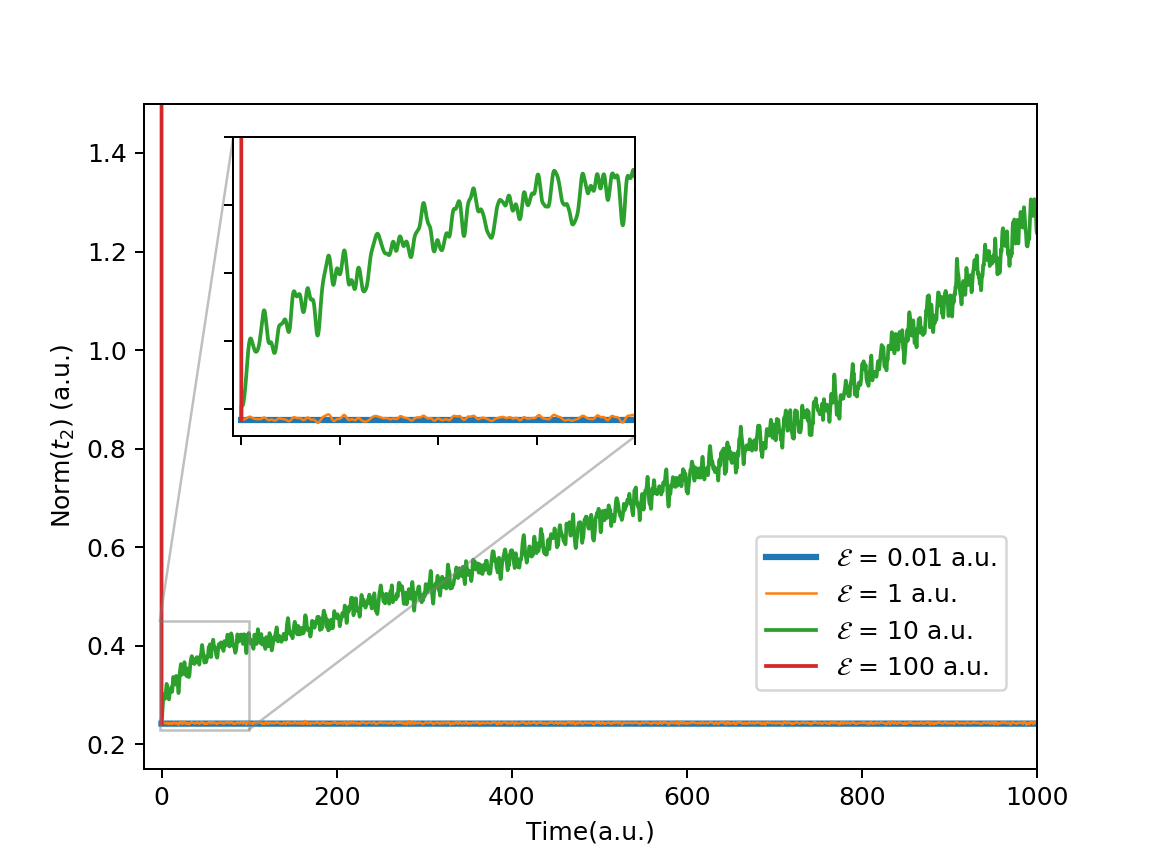}
    \caption{Comparison of the norm of $\hat{T}_2$ amplitudes during an
RT-CC/cc-pVDZ simulation for H$_2$O using a short Gaussian pulse and the RK4
integrator for different field strengths, ${\cal E}$.}
    \label{fig:t2-mag}
\end{figure}

For strong external fields, the numerical stability of the propagation
becomes challenging because the wave function amplitudes fluctuate rapidly
leading to a large local error.  For example, Fig.~\ref{fig:t2-mag} depicts
the norm of the $\hat{T}_2$ amplitudes from an RT-CCSD/cc-pVDZ simulation
of our H$_2$O molecule at several different field strengths of the Gaussian
pulse in Eq.~(\ref{eq:field}) across a 1000 a.u.\ propagation using the RK4
integrator and a step size of $h = 10^{-2}$ a.u. On the scale of the
figure, the weaker fields of ${\cal E} = 0.01$ and $1.0$ a.u.\ induce
relatively small fluctuations in the amplitudes, while a $10.0$ a.u.\ field
yields much larger oscillations.  When the field strength is increased to
${\cal E} = 100.0$ a.u.\, the fluctuations are so great that the
propagation diverges.  In such cases, the TDSE is commonly referred to as a
``stiff equation'' in that the chosen step size must be extremely small to
maintain the stability of the propagation.  For our ${\cal E} = 100.0$
a.u.\ test case, we find that a step size of $h = 10^{-5}$ a.u.\ is
necessary to maintain the integrity of the simulation, which is clearly
impractical for realistic applications.

For the strong-field simulation, it is noteworthy that the divergence
begins near the peak of the of the Gaussian pulse, suggesting that one
might need only decrease the step size while the field is on (the ``bumpy''
portion of the trajectory) and shift to larger values once the field has
decayed.  We therefore carried out a test simulation using $h_1 = 10^{-5}$
a.u.\ when the field is non-zero and $h_2 = 0.01$ a.u.\ otherwise.  The
overhead of this approach is obviously the number of additional residual
evaluations necessary during the pulse, $\frac{\Delta t}{h_{1}} -
\frac{\Delta t}{h_2}$, where $\Delta t$ is the duration of the
field.  The goal is to use steps from $t_{0}$ to $t_{f}$ to track the
interaction with the field precisely, while still minimizing the
computational cost for the overall propagation. 

\begin{figure}[htbp]
    \centering
    \begin{subfigure}[H]{0.475\textwidth}
        \centering
        \includegraphics[width=\textwidth]{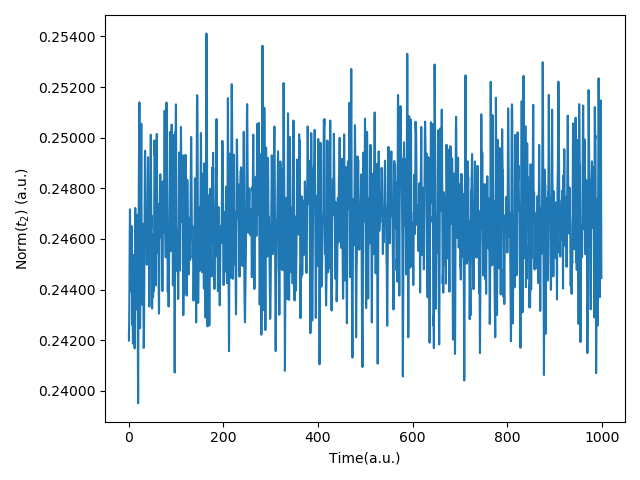}
    \end{subfigure}
    \hfill
    \begin{subfigure}[H]{0.475\textwidth}
        \centering
        \includegraphics[width=\textwidth]{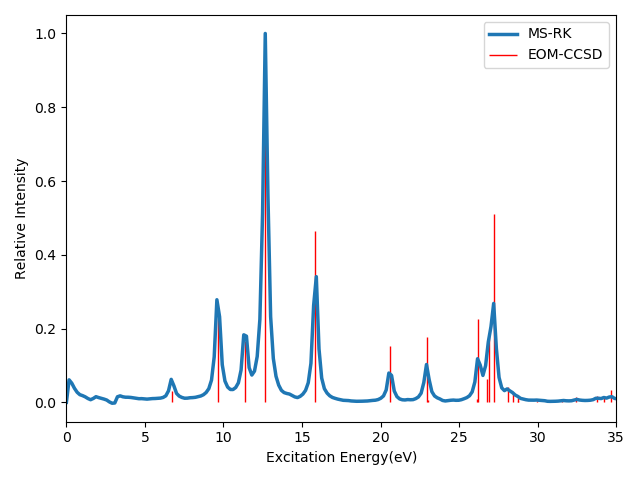}
    \end{subfigure}
    \\
    \begin{subfigure}[H]{0.475\textwidth}
        \centering
        \includegraphics[width=\textwidth]{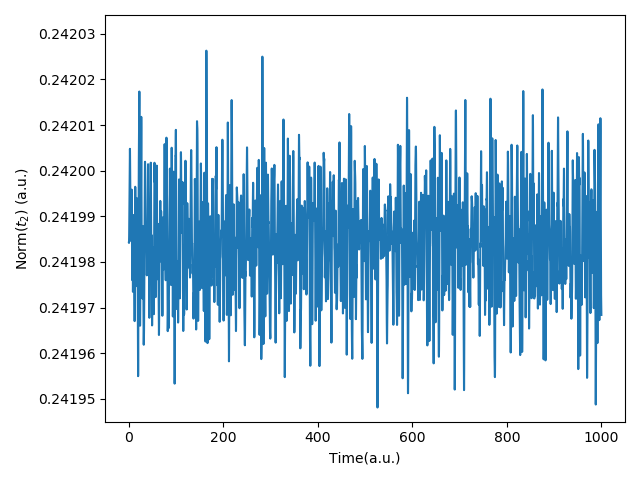}
    \end{subfigure}
    \hfill
    \begin{subfigure}[H]{0.475\textwidth}
        \centering
        \includegraphics[width=\textwidth]{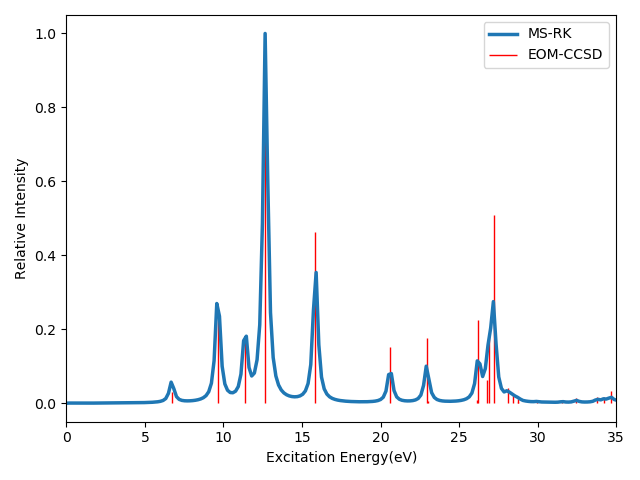}
    \end{subfigure}
    \caption{A $t=1000$ a.u.\ simulation of H$_2$O in the presence of a
    strong ${\cal E} = 100.0$ a.u.\ field with a width of $10^{-4}$ a.u.\
    (upper plots) and $10^{-6}$ a.u.\ (lower plots) at the RT-CCSD/cc-pVDZ
    level of theory using a mixed time-step RK4 approach.  The norm of the
    $\hat{T}_2$ amplitudes are depicted in the left-hand plots, and the
    absorption spectrum is shown in the right-hand plots.}
\label{fig:ms-results}
\end{figure}

In order to test this approach, we chose a very narrow Gaussian pulse with
$\nu=0.0005$ a.u.\ and $\sigma=0.0001$ a.u.\ according to the magnitude of
$h_{1}$, once again for an RT-CCSD/cc-pVDZ calculation for a single H$_2$O
molecule, but for a 1000 a.u.\ propagation.  For ${\cal E} = 100.0$ a.u.\
we found that the proposed mixed-step-size approach recovered a stable
propagation, as shown in Fig.~\ref{fig:ms-results}, whose upper-left-hand
plot depicts the norm of the $\hat{T}_2$ wave function parameters as a
function of time. The amplitude of the oscillation is clearly greater than
that induced by weaker fields, which is expected, but it still does not
diverge throughout the propagation. The corresponding absorption spectrum
in the lower-right-hand plot of Fig.~\ref{fig:ms-results} is nearly the
same as that produced by a weaker field in Fig.~\ref{fig:ck-results}, apart
from some additional noise in the low-frequency regime.

The low-frequency noise disappears, however, if we also choose an even
narrower Gaussian pulse (so as to approximate a strong-field Dirac-delta
pulse) in conjunction with the mixed step-size approach described above.
To demonstrate this we chose parameters for the external field to be
$\cal{E} = 100$ a.u., $\sigma = 10^{-6}$ a.u. and $\nu = 5\times10^{-6}$
a.u.\ with a corresponding step size of $h_{1} =10^{-7}$ a.u.\ during the
pulse and the usual $h_{2} =0.01$ a.u.\ is used after field is off. (Note
that we carried out this calculation in single-precision, and thus
$10^{-7}$ is the smallest scale that can be selected to retain the required
accuracy.) As shown in the lower plots of Fig.~\ref{fig:ms-results}, the
even smaller step size (compared to $h=10^{-5}$ a.u.) gives rise to a more
stable propagation, and therefore, a higher-quality spectrum that avoids
the extra noise in the low-frequency range that appears in the upper-right
panel of Fig.~\ref{fig:ms-results}.  It should also be noted that the
overhead of this calculation is the same as the one with a wider Gaussian
pulse since the ratio $\frac{\Delta t}{h_{1}}$ is unchanged.  Furthermore,
while a mixed step-size approach will necessarily incur more overhead for
more comples field shapes, the stability offered by the algorithm may be
worth the additional expense. Finally we note that, while further testing is necessary to determine the robustness of this
approach across a range of molecular systems, the propagation is stable even for the same molecule and the aug-cc-pVDZ
basis set and with the oxygen $1s$-core electrons included in the correlation treatment.

%% file: conc.tex
\section{Conclusions} \label{conc}

In this work, we have explored several approaches for improving the
efficiency of the real-time coupled cluster singles and doubles method.
Through a number of numerical experiments on absorption spectra of small
clusters of water molecules, we have found that lowering the arithmetic
representation of the wave function from double- to single-precision yields
negligible differences in the resulting spectra, but speeds up the
calculations by nearly a factor of two compared to the conventional
double-precision implementation.  We have additionally found that migration
of the data and the corresponding tensor contractions from CPU to GPU
utilizing the the PyTorch framework produces a further overall speedup of a
factor of 14.  Based on the rapidly growing computational power of GPU
hardware and their supporting software ecosystem, we intend to carry out
further investigation and optimization of our GPU implementation for
calculations on larger molecular systems. 

We have also investigated a variety of numerical integration schemes for
improving the stability and efficiency of the RT-CC approach, especially
focusing on adaptive integrators that can adjust the step size during the
time-propagation.  In particular, we demonstrate that the Cash-Karp
integrator, which uses an estimate of the local error in each iteration, can
accordingly adjust the time-step to optimize the simulation in terms of
both computing time and numerical stability.  However, for very strong
external fields, such as a narrow Gaussian envelope or delta-pulse, both of
which are commonly used in such simulations, we find that even a
straightforward mixed-step integrator based on the fourth-order Runge-Kutta
algorithm is capable of providing a stable propagation provided a small
enough step size is used in for the duration of the field.  Such an
algorithm should be quite favorable for such calculations with intense, but
narrow laser pulses since it enables the existing RT-CCSD method to be
generally used without any substantial modification to the algorithm or
excessively increased computational cost. While further tests are required to determine the generality and robustness of
this approach, the current results are encouraging.